\documentclass[aps,twocolumn]{revtex4}
\usepackage{graphicx}
\usepackage{bm}

\begin{document} 

\title{\bf The thermodynamic dual structure of linear-dissipative
  driven systems} 

\author{Eric Smith}
\affiliation{Santa Fe Institute, 1399 Hyde Park Road, Santa Fe, NM 87501}

\date{\today}
\begin{abstract}

The spontaneous emergence of dynamical order, such as persistent
currents, is sometimes argued to require principles beyond the entropy
maximization of the second law of thermodynamics.  I show that, for
linear dissipation in the Onsager regime, current formation can be
driven by exactly the Jaynesian principle of entropy maximization,
suitably formulated for extended systems and nonequilibrium boundary
conditions.  The Legendre dual structure of equilibrium thermodynamics
is also preserved, though it requires the admission of current-valued
state variables, and their correct incorporation in the entropy. 

\end{abstract}

\maketitle

The search for a theory of the emergent order found in many driven
systems, whether ``self-organization'' or ``dissipative
structures''~\cite{selforg,Prig}, presumes that these systems are not
adequately described by statistical principles such as the second law
of thermodynamics.  In a narrow sense this is certainly true, as all
driven dissipative systems conduct energy and possibly matter between
dissimilar reservoirs.  However, the second law of thermodynamics is
essentially an {\em informational} principle~\cite{Jaynes}, that
stable macrostates correspond to maximally disordered distributions of
microstates, constrained by average values of dynamically conserved
quantities like energy, with Shannon-Boltzmann entropy the measure of
disorder.  Nothing precludes the application of this principle to
time-dependent systems.

A phenomenological rule advanced to explain the form of currents in
near-equilibrium systems is maximal entropy {\em
  production}~\cite{Onsager,Prig}, where ``entropy'' means the usual
coarse-grained equilibrium functional of time-reversal-invariant state
variables.  Entropy production is often perceived as independent from
the second law, capable of opposing it and favoring the emergence of
``order'' in some settings.  However, the entropy production function
presumes equilibrium state variables remain meaningful, implying that
the fine-grained configuration of the system locally obeys the second
law in the usual sense.

Here I consider a model of linear dissipation in the Onsager
regime~\cite{Onsager}, obtained from first principles by typical
methods of quantum optics~\cite{decoherence}, but given sufficient
structure to represent an {\em extended} system capable of coupling to
multiple reservoirs with dissimilar temperatures.  In equilibrium
systems, the value of any intensive state variable is necessarily the
same for all system degrees of freedom, and dual to each intensive
variable is an extensive distribution-averaged observable obtained by
Legendre transformation.  The imposition by the boundary conditions of
multiple values for a single intensive variable overconstrains the
system relative to an equilibrium, and leads to spontaneous emergence
of persistent currents.  However, I show that the stable states of
such driven systems remain consistent with Jaynesian maximization of
the exact entropy, subject to constraint of distribution-averaged
observables.  Further the Legendre dual structure of equilibrium
thermodynamics is preserved, if one admits current-valued
(time-reversal asymmetric) as well as charge-valued (symmetric) state
variables, and computes the entropy as a function of both.

The result is that macroscopic as well as microscopic features of the
distribution are determined uniformly by entropy maximization.  The
dynamical distribution is more ordered than an equilibrium
distribution at the same average energy simply because its boundary
constraints are more structured, which is also the origin of the
second law in equilibrium~\cite{MurraySeth}.  Maximum production of
coarse-grained entropy may be recovered as a phenomenological rule
from fine-grained entropy maximization at each time, once the latter
has ensured that all local marginal distributions are thermal and that
the equilibrium state variables have their usual meaning.



The model is similar to Fock space models used in quantum optics,
where linear dissipation is readily derived as a universality
class~\cite{decoherence}.  The physical state space of the system
results from the action of $D$ orthogonal creation and annihilation
operators starting from a ground (ket) state $\left| 0 \right>$.  They
have commutation relations $\left[ a^{\mu} , a^{\dagger}_{\nu} \right]
= {\delta}_{\nu}^{\mu}$, $\mu , \nu \in 1 , \ldots , D$.  The system
has a ``spatial'' basis indexed $i$, in which reservoir coupling is
diagonal, and a (generally different) basis indexed $\mu$ in which the
Hamiltonian is diagonal.

In any basis, the diagonal elements of the dyadic matrix operator
$\hat{n} \equiv a^{\dagger} a$ (i.e., $\left[ {\hat{n}}_{\nu}^{\mu}
\right] \equiv \left[ a^{\dagger}_{\nu} a^{\mu} \right]$) constitute a
set of the independent number components in the Fock space.  The
Hamiltonian (up to constants) is written $\hat{H} \equiv \mbox{Tr}
\left[ E \hat{n} \right]$, with eigenvalues $E^{\nu}_{\mu} \equiv
{\delta}^{\nu}_{\mu} E_{\left( \mu \right)}$.

For a column vector $\xi \equiv \left[ {\xi}^{\mu} \right]$ of complex
scalars, a general coherent state for the system is compactly written
\begin{equation}
  \left| \xi \right> \equiv
  e^{ - 
    \left( {\xi}^{\dagger} \xi \right) /2
  }
  \sum_{N = 0}^{\infty}
  \frac{
    {
      \left(
        a^{\dagger} \cdot \xi
      \right)
    }^N
  }{
    N !
  }
  \left| 0 \right> . 
\label{eq:vec_coh_state_def}
\end{equation}
A subset of density matrices in the Glauber-Sudarshan
$P$-representation~\cite{decoherence}, which represent thermal and
near-thermal conditions, are those diagonal in $\left| \xi \right>$
with Gaussian density
\begin{equation}
  \rho = 
  \mbox{Det} \! 
  \left( K \right)
  \int 
  \frac{
    d {\xi}^{\dagger} d \xi
  }{
    {\pi}^D
  } \, 
  e^{
    - {\xi}^{\dagger} K \xi
  } \, 
  \left| \xi \right>
  \left< \xi \right| ,
\label{eq:density_Gauss_coherent}
\end{equation}
where $K$ is a $D \times D$ Hermitian kernel.  Gaussian-coherent
densities~(\ref{eq:density_Gauss_coherent}) have the property that
all marginal distributions for a single degree of freedom
$ P_{\alpha} \! \left( n_{\alpha} \right) \equiv 
  \mbox{Tr} 
  \left( 
    \rho  
    \left| n_{\alpha} \right>
    \left< n_{\alpha} \right|
  \right)$ are exponential in
$n_{\alpha}$ in any basis~\cite{refrig}.
They include the equilibrium thermal distributions ($ K^{\mu}_{\nu}
\equiv {\delta}^{\mu}_{\nu} \left( e^{\beta E_{\left( \mu \right)}} -1
\right) $), and also general time-dependent distributions created by
linear dissipation, from arbitrary Gaussian-coherent initial
conditions.  Though all marginals are effectively thermal, in
nonequilibrium cases different marginals generally have different
effective temperatures.

The kernel $K$ is the inverse expected number operator
\begin{equation}
  n \equiv 
  \mbox{Tr} \left( \rho \hat{n} \right) = 
  K^{-1} , 
\label{eq:bar_n_mat_def}
\end{equation}
and the exact quantum entropy has the von Neumann form~\cite{refrig}
\begin{eqnarray}
  S \left( \rho \right) 
& \equiv &  
  - \mbox{Tr}
  \left(
    \rho \log \rho
  \right) 
\nonumber \\
& = & 
  \mbox{Tr}
  \left[
    \left( I + n \right) 
    \log \left( I + n \right) - 
    n \log n
  \right] , 
\label{eq:S_GCE_def_eval}
\end{eqnarray}
(in which parentheses denote density matrix trace, and square brackets
the scalar matrix trace over index $\mu$).  Gaussian-coherent
ensembles are specified among general density matrices $\rho$ by
standard Jaynesian entropy maximization, 
\begin{equation}
  {\delta}_{\rho, \lambda} 
  \left\{
    S \left( \rho \right) - 
    \mbox{Tr} \left[ 
      \lambda
      \left(
        \mbox{Tr} \left( \rho \hat{n} \right) - K^{-1}
      \right)
    \right]
  \right\} = 0 , 
\label{eq:max_ent_trace}
\end{equation}
where $\lambda$ is a matrix of Lagrange multipliers enforcing a
constraint on the {\em trace} of the number operator $\hat{n}$.  The
partition function of the
distribution~(\ref{eq:density_Gauss_coherent}) is indistinguishable
from that for a tensor product of thermal states (diagonalize $K$),
and evaluates to
\begin{equation}
  \log Z = 
  \mbox{Tr} 
  \left[
    \log 
    \left( 
      I + n
    \right)
  \right] = 
  S \! \left( n \right) - 
  \mbox{Tr} 
  \left[
    n
    \frac{
      \delta S \! \left( n \right)
    }{
      \delta n
    }
  \right] . 
\label{eq:log_Z_expand}
\end{equation}

The master equation for a general system density $\rho$, minimally
coupled to as many as $D$ reservoirs (one per system degree of
freedom), is~\cite{decoherence}
\begin{eqnarray}
\lefteqn{
  \frac{\partial \rho}{\partial t} = 
  i \left[ 
    \hat{H} , \rho
  \right] 
} & & 
\nonumber \\
& & 
  \mbox{} - 
  \frac{r}{2}
  \left\{ 
    \hat{N} , \rho
  \right\} + 
  r 
  a^{\mu} \rho a^{\dagger}_{\mu} 
\nonumber \\
& & 
  \mbox{} - 
  r 
  \left\{ 
    {\hat{\Sigma}}_R , \rho
  \right\} - 
  r 
  \mbox{Tr}
  \left( n_R \right) \rho + 
  r 
  {\left( n_R \right)}^{\nu}_{\mu}
  \left[ 
    a^{\dagger}_{\nu} \rho a^{\mu} + 
    a^{\mu} \rho a^{\dagger}_{\nu} 
  \right] . 
\nonumber \\
& & 
\label{eq:states_Ops_evolve}
\end{eqnarray}
$\hat{N} \equiv \mbox{Tr} \left[ \hat{n} \right] = a^{\dagger}_{\mu}
a^{\mu}$, is the total number operator, and ${\hat{\Sigma}}_R \equiv
\mbox{Tr} \left( \hat{n} n_R \right) = a^{\dagger}_{\nu} {\left( n_R
\right)}^{\nu}_{\mu} a^{\mu}$ is a source of particles from the
reservoirs.  The detailed structure of the reservoirs does not matter,
though $n_R$ has the interpretation of a mean excitation number in
reservoirs that are also systems of linear oscillators.  The
entropy~(\ref{eq:S_GCE_def_eval}) is coarse-grained~\cite{MurraySeth}
because it results from projection of a reversible $\left(
\mbox{system} \otimes \mbox{reservoir} \right)$ ensemble onto the
degrees of freedom of the system alone.  Constant $n_R$ characterizes
sufficiently high temperatures and reservoir
dimensionality~\cite{decoherence}.  Finally,
Eq.~(\ref{eq:states_Ops_evolve}) is an effective field equation, in
which Green's function corrections from system-reservoir coupling have
been absorbed in specifying $\hat{H}$ and ${\hat{\Sigma}}_R$.

Eq.~(\ref{eq:states_Ops_evolve}) preserves Gaussian-coherent
ensembles, with only mean particle number evolving as 
\begin{equation}
  \frac{dn}{dt} = 
  i 
  \left[
    E , n
  \right] + r 
  \left(
    n_R - n 
  \right)
\label{eq:evolve_n_relation_R}
\end{equation}
Particles diffuse from the system in proportion to $n$, and conversely
from the reservoir in proportion to $n_R$.  In a high-temperature
(Boltzmann) regime, particle statistics play no special role, and
linear particle exchange is expected to be representative of ubiquitous
diffusion relations such as the Fourier and Ohm laws.

If $\left[ E , n_R \right] = 0$ the eigenvectors of $\hat{H}$ are
coupled independently to different reservoir components, and in steady
state $\rho$ decomposes into a tensor product of independent thermal
subsystems.  $\left[ E , n_R \right] \neq 0$ describes ``open''
systems, in which unequal intensive state variables from different
components of $n_R$ induce persistent currents at steady state.

The entropy~(\ref{eq:S_GCE_def_eval}) evolves under
Eq.~(\ref{eq:evolve_n_relation_R}) as
\begin{eqnarray}
  \frac{
    d
  }{
    dt
  } 
  S \! \left( \rho \right)
& = & 
  -r 
  \mbox{Tr}
  \left[
    \left( n  - n_R \right)
    \frac{\delta S}{\delta n}
  \right] = 
  r 
  \mbox{Tr} 
  \left[ 
    \left( \rho  - {\rho}_R \right)
    \log \rho
  \right] 
\nonumber \\
& = & 
  r
  \left[
    \Delta \! 
    \left( \rho ; {\rho}_R \right) - 
    S \left( \rho \right) + 
    S \left( {\rho}_R \right) 
  \right] , 
\label{eq:S_dist_from_cg}
\end{eqnarray}
where $\Delta \! \left( \rho ; {\rho}_R \right)$ is the
Kullback-Leibler pseudodistance~\cite{CoverThomas}
\begin{eqnarray}  
\lefteqn{
  \Delta 
  \left( \rho ; {\rho}_R \right) \equiv 
  \mbox{Tr}
  \left[
    {\rho}_R 
    \left( 
      \log {\rho}_R - 
      \log \rho
    \right)
  \right] 
} & &   
\nonumber \\
& = & 
  \mbox{Tr}
  \left[
    \left( I + n_R \right) 
    \log \left( I + n \right) 
    { \left( I + n_R \right) }^{-1} - 
    n_R \log n n_R^{-1} 
  \right] , 
\nonumber \\
& & 
\label{eq:KL_compute}
\end{eqnarray}
and ${\rho}_R$ is to be understood as the Gaussian-coherent ensemble
the reservoirs ``attempt to impose'' through $n_R$.  

The steady state condition $dS / dt = 0$ is equivalent to the
condition $S \! \left( \rho \right) - S \! \left( {\rho}_R \right) =
\Delta \! \left( \rho ; {\rho}_R \right)$ that $\rho$ be a
coarse-graining of ${\rho}_R$~\cite{MurraySeth}.  Geometrically, $n_R
= n - i \left[ E, n \right] / r$ is in the tangent plane to the
surface $\delta S = 0$ at $n$, because although $S \!  \left( \rho
\right)$ arises from a projection, unitary evolution $i \left[ E , n
\right]$ within the system preserves the value of $S \!  \left( n
\right)$.

We will generally retain time-dependent $n$ for comparison to the
Onsager construction, but it is convenient to have a closed form for
the asymptotic late-time distribution, denoted 
\begin{equation}
  \bar{n} = 
  \int_0^{\infty}
  r \, dt \, 
  e^{-rt}
  e^{
    i Et
  }
  n_R
  e^{
    - i Et 
  } . 
\label{eq:SS_asympt_repn}
\end{equation}
In components in the Eigenbasis of the Hamiltonian,
\begin{equation}
  {\left( \bar{n} \right)}_{\mu}^{\nu} = 
  \frac{
    {
      \left( n_R \right)
    }_{\mu}^{\nu}
  }{
      1 - i 
      \left( 
        E_{\left( \mu \right)} - E_{\left( \nu \right)}
      \right) / 
      r
  } . 
\label{eq:E_eigs_n_sol}
\end{equation}

\begin{figure}[ht]
  \begin{center} 
  \includegraphics[scale=0.6]{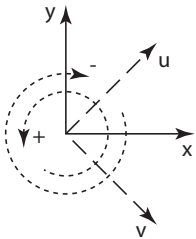}
  \caption{
  Three bases for the two-dimensional oscillator.  $xy$ (solid) is the
  eigenbasis of the Hamiltonian.  $uv$ (dash) is the ``spatial'' basis
  in which a thermal asymmetry is imposed by $n_R$.  $\pm$ (dotted) is
  the current basis excited spontaneously, and switched under time
  reversal. 
    \label{fig:SU2_basis} 
  }
  \end{center}
\end{figure}

The physical reason for the emergence of currents from coarse-graining
is nicely illustrated in the asymptotic solution to the simplest
nontrivial example, a two-dimensional oscillator with Hamiltonian
\begin{equation}
  E = 
  \left[
    \begin{array}{cc}
      E_x &     \\
          & E_y 
    \end{array}
  \right] , 
\label{eq:E_SU2_diag}
\end{equation}
and $E_x \neq E_y$.  General Hermitian $n$ take the form
\begin{equation}
  n = 
  \left[
    \begin{array}{cc}
      n_0 + n_3   & n_1 + i n_2 \\
      n_1 - i n_2 & n_0 - n_3 
    \end{array}
  \right] , 
\label{eq:n_SU2_diag}
\end{equation}
with coefficients related to the physical state basis of
Fig.~\ref{fig:SU2_basis} by $2 n_3 = n_x - n_y$, $2 n_1 = n_u - n_v$,
$2 n_2 = n_{+} - n_{-}$.  If by convention $n_{\left( x,y \right)}$
and $n_{\left( u,v \right)}$ refer to standing waves, $n_{\pm}$ are
traveling waves exchanged under time reversal.  Then the coefficients
of real $n$ are charge-valued state variables, and the single
imaginary coefficient is a current-valued state variable.  As the
entropy~(\ref{eq:S_GCE_def_eval}) is preserved by arbitrary
similarity transformation of $n$ under $\mbox{SU} \!  \left( 2
\right)$, charge and current state variables have identical
interpretations in terms of statistical uncertainty.

\begin{figure}[ht]
  \begin{center} 
  \includegraphics[scale=0.5]{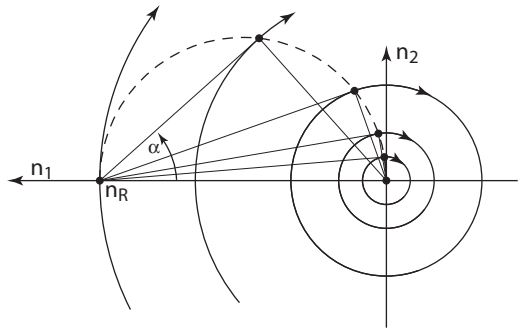}
  \caption{
  Coarse-graining from a charge $n_1$ to a current $n_2$.  Concentric
  circles are Hamiltonian orbits, and dashed circle is the set of
  coarse grainings from given ${\left( n_R \right)}_1$.  The angle
  $\alpha$ obeys $\tan \alpha = 1 / \lambda = r / \left( E_x - E_y
  \right)$.
    \label{fig:SU2_coarsegrain} 
  }
  \end{center}
\end{figure}

Hamiltonian evolution preserves $n_0$ and $n_3$, and oscillates net
charge and current excesses in time as
\begin{equation}
  {
    \left( n_1 + i n_2 \right)
  }_t = 
  e^{
    i \left( E_x - E_y \right) t 
  }
  {
    \left( n_1 + i n_2 \right)
  }_0 . 
\label{eq:Hamiltonian_n1n2}
\end{equation}
The set of coarse-grainings of real $n_R$ preserving $n_0$ and $n_3$
is given by ${\left( n_1 + i n_2 \right)}_{\lambda} = {\left( n_R
\right)}_1 / \left( 1 - i \lambda \right)$ for real $\lambda$, and
shown in Fig.~\ref{fig:SU2_coarsegrain}.  By
Eq.~(\ref{eq:E_eigs_n_sol}) for $\bar{n}$, we solve for $\lambda =
\left( E_x - E_y \right) / r$. The current magnitude $\left|
{\bar{n}}_2 \right|$ is maximized at $r^2 = {\left( E_x - E_y
\right)}^2$.

We may generate spontaneous currents in an extended system as follows:
At equilibrium $n_0 \pm n_3 \neq 0$, while $n_1 = 0$ (reflection
symmetry), and $n_2 = 0$ (time-reversal invariance).  If we couple the
orthogonal $\left( u, v \right)$ operators to static reservoirs with
different temperatures, consistent with the same $n_0 \pm n_3$, linear
dissipation attracts $n$ in Eq.~(\ref{eq:evolve_n_relation_R}) toward
a charge asymmetry ${\left( n_R \right)}_1 \neq 0$, which refines the
equilibrium ensemble with a new constraint, thus reducing its entropy.
Dissipation just balances Hamiltonian charge/current conversion at a
coarse-graining $\bar{n}$ of $n_R$, where $0 < \left| {\bar{n}}_1 + i
{\bar{n}}_2 \right| < \left| {\left( n_R \right)}_1 \right|$, allowing
a larger entropy than $S \! \left( n_R \right)$ but less than
equilibrium for the same average system energy.


Returning to the general case, we may derive the classical
thermodynamic dual structure for arbitrary target $n_{\tau}$.  If through
Eq.~(\ref{eq:SS_asympt_repn}) we define
\begin{equation}
  {\Lambda}_{\tau} \equiv
  \frac{
    \delta S \! \left( n_{\tau} \right)
  }{
    \delta n_{\tau}
  } = 
  \log 
  \left( 
    1 + {n}^{-1}_{\tau}
  \right) , 
\label{eq:vary_nbar_A}
\end{equation}
the remaining degrees of freedom in arbitrary $\mbox{Tr} \left( \rho
\hat{n} \right)$ are fixed by maximizing $S \! \left( \rho \right)$
subject to the single constraint $\mbox{Tr} \left[ \left( \mbox{Tr}
\left( \rho \hat{n} \right) - n_{\tau} \right) {\Lambda}_{\tau} \right] = 0$,
because the set $\delta S \ge 0$ from any $n_{\tau}$ is convex.  Entropy
maximization with one Lagrange multiplier for an extensive trace
constraint,
\begin{equation}
  {\delta}_{\rho ,\lambda} 
  \left\{
    S \! \left( \rho \right) - 
    {\lambda}_{\tau}
    \left[
      \mbox{Tr} 
      \left(
        \mbox{Tr} 
        \left[
          \hat{n}
          {\Lambda}_{\tau}
        \right] 
        \rho 
      \right) - 
      \mbox{Tr} 
      \left[
        n_{\tau} {\Lambda}_{\tau}
      \right] 
    \right]
  \right\} = 0 ,  
\label{eq:max_S_rho_oneparm}
\end{equation}
is thus a {\em sufficient} principle to extract $\rho \! \left( n_{\tau}
\right)$, but fails to capture the physical role of the temperatures
in the reservoir as independently specified intensive state variables.

The state variable interpretation of reservoir parameters is recovered
naturally, and indeed linearly, in the Onsager regime of high
temperature and weak perturbation away from equilibrium.  Suppose, for
example, that the trace constraint on $n_R$ comes from average energy
and some other charge matrix $Q$.  Suppose a high temperature
$\bar{\beta} E_{\left( \mu \right)} \ll 1$, and weak perturbation
${\beta}^{\prime} Q_{\left( \nu \right)} / \bar{\beta} E_{\left( \mu
\right)} \ll 1$, for all eigenvalues $E_{\left( \mu \right)}$ of $E$
and $Q_{\left( \nu \right)}$ of $Q$.  $Q$ is a charge if it is
Hermitian with real coefficients.  With these trace constraints, we
may expand $n_R$ to leading order
\begin{equation}
  n_R = 
  {
    \left( 
      e^{
        \bar{\beta} E + {\beta}^{\prime} Q 
      } - 
      1 
    \right)
  }^{-1} \approx 
  \frac{E^{-1}}{\bar{\beta}} - 
  \frac{E^{-1}}{\bar{\beta}}
  {\beta}^{\prime} Q 
  \frac{E^{-1}}{\bar{\beta}} . 
\label{eq:n_R_simpleform}
\end{equation}
Under Eq.~(\ref{eq:evolve_n_relation_R}) we may then write 
\begin{equation}
  n \approx 
  \frac{E^{-1}}{\bar{\beta}} - 
  \frac{E^{-1}}{\bar{\beta}}
  {\beta}^{\prime} J 
  \frac{E^{-1}}{\bar{\beta}} \approx 
  {
    \left( 
      e^{
        \bar{\beta} E + {\beta}^{\prime} J 
      } - 
      1 
    \right)
  }^{-1} , 
\label{eq:barn_simpleform}
\end{equation}
as long as 
\begin{equation}
  \frac{dJ}{dt} = 
  i \left[ E , J \right] + 
  r \left( Q - J \right) . 
\label{eq:J_evol_c_num}
\end{equation}
If we think of $Q$ as denominated in energy units, $1 / \left(
\bar{\beta} \mp {\beta}^{\prime} \right) \equiv kT_{\pm}$ are the two
temperature parameters represented in the environment.  Dual to the
average temperature $1 / \bar{\beta}$ is an average energy
\begin{equation}
  \mbox{Tr}
  \left[
    \bar{n} E
  \right] = 
  \mbox{Tr}
  \left[
    n_R E
  \right] \equiv 
  \bar{\cal E} , 
\label{eq:conserved_trace}
\end{equation}
(equal to that in $n_R$), and dual to ${\beta}^{\prime} /
{\bar{\beta}}^2$ is a new trace
\begin{equation}
  \mbox{Tr}
  \left[
    n J
  \right] \equiv 
  {\cal J} . 
\label{eq:J_trace}
\end{equation}
Denoting the Hermitian operators formed from $Q$ and $J$ 
$\hat{Q} \equiv \mbox{Tr} \left[ Q \hat{n} \right]$ and 
$\hat{J} \equiv \mbox{Tr} \left[ J \hat{n} \right]$ respectively, the
steady state distribution solves the maximization problem
\begin{equation}
  {\delta}_{\rho , \lambda , {\lambda}^{\prime}}
  \left\{
    S \! \left( \rho \right) - 
    \lambda 
    \left[
    \mbox{Tr} 
    \left(
      \hat{H}
      \rho
    \right) - 
    \bar{\cal E}
    \right] - 
    {\lambda}^{\prime} 
    \left[
    \mbox{Tr} 
    \left(
      \hat{J}
      \rho
    \right) - 
    {\cal J}
    \right]
  \right\} = 0 . 
\label{eq:max_S_rho_thermal}
\end{equation}
Dual to entropy maximization~(\ref{eq:max_S_rho_thermal}) under
extensive state variable constraints is the minimization with
intensive constraints, 
\begin{equation}
  {\delta}_{\rho}
  \left\{
    \bar{\beta}
    \mbox{Tr} 
    \left(
      \hat{H}
      \rho
    \right) + 
    {\beta}^{\prime} 
    \mbox{Tr} 
    \left(
      \hat{J}
      \rho
    \right) - 
    S \! \left( \rho \right)
  \right\} = 0 . 
\label{eq:min_log_Z_general}
\end{equation}
Reducing to the residual dependence on $n$, we recognize that this is
the minimization of the log inverse partition
function~(\ref{eq:log_Z_expand})
\begin{equation}
  {\delta}_{n}
  \left\{
    - \log Z
  \right\} = 
  {\delta}_{n}
  \left\{
    \mbox{Tr} 
    \left[
      \left(
        \bar{\beta} E + 
        {\beta}^{\prime} J 
      \right)
      n 
    \right] - 
    S \! \left( n \right)
  \right\} = 0 , 
\label{eq:min_log_Z_general_n}
\end{equation}
equivalent to minimization of the Helmholtz free energy for the
equilibrium canonical ensemble. 

The phenomenological result of maximal entropy production, a
variational principle for the diffusive currents of equilibrium state
variables, follows from minimization of the generalized exact free
energies~(\ref{eq:min_log_Z_general_n}) in the Onsager regime.  The
phenomenological entropy is obtained by coarse-graining the exact
distribution to a product of its marginals on different spatial
positions.  The equivalent operation here is the projection of $\rho$
onto its diagonal components $\tilde{\rho}$ in a Fock space over
independent number excitations.  The resulting coarse-grained entropy
is a function of the charge-valued state variables only.

As we have only introduced notation and derived linear particle
exchange for factoring the $\left( \mbox{system} \otimes
\mbox{reservoir} \right)$ distribution into independent marginals,
that partition will be used as an example.  It simplifies the
presentation to consider the reservoir a unified system in its own
right (rather than a set of independent components), with state
variable $n_R$, and distribution ${\rho}_R$ extremized as in
Eq.~(\ref{eq:min_log_Z_general_n}), except with trace constraint
$\bar{\beta} E + {\beta}^{\prime} Q$, for consistency with
Eq.~(\ref{eq:n_R_simpleform}).  Label the reservoir partition function
$Z_R$, and its entropy $S_R$ computed from ${\rho}_R$ as for $S \!
\left( \rho \right)$.  Let overdot denote time differentiation.  Then
consider the potential 
\begin{eqnarray}
\lefteqn{
  \frac{d}{dt}
  \log \left( Z Z_R \right) = 
  \dot{S} - 
  \mbox{Tr} 
  \left[
    \left(
      \bar{\beta} E + 
      {\beta}^{\prime} J 
    \right)
    \dot{n} - 
    {\beta}^{\prime} \dot{J} n 
  \right] 
} & & 
\nonumber \\
&  &
  \mbox{} +  
  {\dot{S}}_R - 
  \mbox{Tr} 
  \left[
    \left(
      \bar{\beta} E + 
      {\beta}^{\prime} Q
    \right)
    {\dot{n}}_R - 
    {\beta}^{\prime} \dot{Q} n_R 
  \right] 
\label{eq:log_ZZR_timeder}
\end{eqnarray}
Use Eq.~(\ref{eq:J_evol_c_num}) to evaluate $\dot{J}$, and by
symmetric treatment of system and reservoir let $\dot{Q} = r \left( J
- Q \right)$ (taking the reservoir energy diagonal in $n_R$ purely for
convenience).  In terms of currents defined phenomenologically from
$n$ and $n_R$, 
\begin{equation}
  j \equiv \dot{n} - i \left[ E , n \right]
\label{eq:n_curr_def}
\end{equation}
and $j_R \equiv {\dot{n}}_R$, we may then write
Eq.~(\ref{eq:log_ZZR_timeder}) as 
\begin{eqnarray}
\lefteqn{
  \frac{d}{dt}
  \log \left( Z Z_R \right) = 
  \mbox{Tr} 
  \left[
    \left( 
      \frac{\delta S}{\delta n} - 
      \bar{\beta} E - 
      {\beta}^{\prime} J 
    \right)
    j 
  \right. 
} & & 
\nonumber \\
&  &
  \mbox{} +  
  \left. 
    \left( 
      \frac{\delta S_R}{\delta n_R} - 
      \bar{\beta} E - 
      {\beta}^{\prime} Q
    \right)
    j_R - 
    {\beta}^{\prime} 
    \left( J - Q \right)
    \left( n - n_R \right)
  \right] . 
\nonumber \\
& & 
\label{eq:log_ZZR_dot_curr_form}
\end{eqnarray}
Maximization of $\log Z$ and $\log Z_R$ at each time is equivalent to
saddle-point extremization of Eq.~(\ref{eq:log_ZZR_dot_curr_form}),
maximizing in $j$ and $j_R$, minimizing in $n$ and $n_R$.  

For high temperatures and linear perturbations it is convenient to
write $n = {\left( \bar{\beta} E \right)}^{-1} \! \! \! + \delta n$,
$n_R = {\left( \bar{\beta} E \right)}^{-1} \! \! \! + \delta n_R$, and
Eq.~(\ref{eq:log_ZZR_dot_curr_form}) is readily expressed to leading
order as a difference of quadratic forms in these variables.
Variation with $\delta n + \delta n_R$ then sets $j + j_R = 0$, and
variation with $j + j_R$ recovers the sum of
forms~(\ref{eq:n_R_simpleform},\ref{eq:barn_simpleform}). Variation
with $\delta n - \delta n_R$ gives $\delta n - \delta n_R$ in terms of
$j - j_R$, recovering the linear dissipation rule.  Evaluating
Eq.~(\ref{eq:log_ZZR_dot_curr_form}) on these three extrema leaves the
function of $\left( j - j_R \right) / 2 \rightarrow j$:
\begin{eqnarray}
\lefteqn{
  \frac{d}{dt}
  \log \left( Z Z_R \right) = 
  4 r 
  \mbox{Tr} 
  \left[
    {\beta}^{\prime}
    \left( Q-J \right)
    j - 
    \frac{1}{2r}
    \left( \bar{\beta} E \right) j 
    \left( \bar{\beta} E \right) j 
  \right] 
} & & 
\nonumber \\
&  &
  \mbox{} +  
  r 
  \left( 2r - 1 \right)
  \mbox{Tr} 
  \left[
    \left( \bar{\beta} E \right) 
    {\beta}^{\prime} \! 
    \left( Q-J \right)
    \left( \bar{\beta} E \right) 
    {\beta}^{\prime} \! 
    \left( Q-J \right)
  \right] . 
\nonumber \\
& & 
\label{eq:log_ZZR_dot_js_form}
\end{eqnarray}
The linear form $\mbox{Tr} \left[ {\beta}^{\prime} \left( Q-J \right)
  j \right]$ is Onsager's ``entropy production rate'', with the
difference of inverse temperatures ${\beta}^{\prime} \! \left( Q-J
\right)$ regarded as constants under variation, while
\begin{equation}
  \phi \! 
  \left( j , j \right) \equiv 
  \frac{1}{2r}
  \mbox{Tr}
  \left[
    \left( \bar{\beta} E \right) j 
    \left( \bar{\beta} E \right) j 
  \right] 
\label{eq:diss_fn_intro}
\end{equation}
is the phenomenological ``dissipation function''.  Under complete
coarse-graining of both system and reservoirs, the single
function~(\ref{eq:log_ZZR_dot_js_form}) would expand into the
structure of physical diffusion currents determined by the
Hamiltonian, though the resulting coarse-grained entropy, even within
the system, would progressively diverge from the exact
value~(\ref{eq:S_GCE_def_eval}).  

This formal demonstration that spontaneous emergence of persistent
currents need not require any new principles beyond Jaynesian entropy
maximization potentially changes our understanding of the statistical
nature of dynamical order.  States with currents are potentially
equivalent to equilibrium ground states, with the dual representation
of nonequilibrium boundary conditions taken up by dynamical or
time-reversal-asymmetric extensive observables.

It was previously shown~\cite{refrig} that entropy maximization
naturally extends from equilibria to cases of thermodynamic
reversibility, an intuitive result because state variables at different
times index the same constraint set.  Less obvious, when such systems
support self-organizing phase transitions~\cite{Carnot1}, their
finite-temperature field theory~\cite{Carnot2} retains the structure
of equilibrium up to analytic continuation, suggesting that nonlinear
positive feedback is also consistent with entropy-maximization, if one
takes care with broken ergodicity.  The current demonstration for
linear-dissipative systems extends this result to cases in which
coarse-grained entropy is not preserved. It remains to determine
whether positive feedbacks that induce phase transitions in
dissipative systems have a similar formulation.

\vspace{12pt}

I am grateful to Insight Venture Partners for support, and to Dave
Bacon and Fred Cooper for most helpful discussions and references.

\end{document}